\documentstyle[12pt]{article}

\begin{document}

\begin{titlepage}
\begin{flushright}
IJS-TP-98/01\\
TECHNION-PH-98-01\\
hep-ph/9801279\\

%4Jan.98\\ 
\end{flushright}

\vspace{.5cm}

\begin{center}
{\Large \bf Long distance contributions  in $D \to V \gamma$ decays\\}

\vspace{.5cm}

{\large \bf S. Fajfer$^{a}$, S. Prelov\v sek$^{a}$ and
P. Singer$^{b}$\\}

\vspace{0.5cm}
{\it a) J. Stefan Institute, Jamova 39, P. O. Box 300, 1001 Ljubljana, 
Slovenia}\vspace{.5cm}

{\it b) Department of Physics, Technion - Israel Institute  of Technology, 
Haifa 32000, Israel\\}

\end{center}

\vspace{1.5cm}

\centerline{\large \bf ABSTRACT}

\vspace{0.5cm}
Using the factorization scheme for the nonleptonic $D \to V V_0$ 
weak amplitudes, we classify all diagrams which 
arise in $D \to V \gamma$ decays and calculate them with the help of the
hybrid model 
which combines  the heavy quark effective theory and the 
chiral Lagrangian approach. 
Thus we determine the long distance contribution to the amplitudes  of 
Cabibbo allowed and Cabibbo suppressed $D \to V\gamma$ decays. 
The calculation of the expected range of the branching ratios 
of nine different $D \to V \gamma$ channels is compared 
with results of other approaches. 
The present work establishes an increase of the parity 
violating contribution in these decays 
in comparison with previous analyses. \\

\end{titlepage}

\setlength {\baselineskip}{0.55truecm}
\setcounter{footnote}{1}    % start footnotes at dagger instead
			     % of '*' (article style only)

\setcounter{footnote}{0}

{\bf 1 Introduction}

\vspace{0.5cm}

The study of nonleptonic decays of charm mesons has been a subject of high 
priority for more than two decades and has resulted in a wealth of 
experimental data, 
which continues to expand at a remarkable rate. This rich source of information 
plays a decisive role in the development of the theoretical treatment of these 
processes, which are governed by the interplay of the weak and 
strong interactions. 

On the other hand, there is very little information available on the sector of 
flavour changing radiative decays of charm mesons, 
in which the electromagnetic interactions is also operative. 

Only some preliminary upper limits, at the $10^{-4}$ range, for branching ratios 
of Cabibbo forbidden decays of 
type $D^0 \to V^0 \gamma$ have been reported so far \cite{S}. 
However, as a result of ongoing efforts \cite{R} it is reasonable to expect that 
the experimental 
data on $D \to V \gamma$ decays will be forthcoming during the next few years. 
The theoretical treatment of these decays must address firstly the 
question of the relative importance of short and long distance contributions. 
For the decays studied here, the short distance process of relevance, 
the $c \to u \gamma$ transition which is driven by the magnetic penguin diagram, 
is exceedingly small being suppressed by GIM cancellation and small 
CKM matrix elements \cite{BGHP}. The inclusion of gluonic corrections 
\cite{BGHP,GHMW} increases the free quark $c \to u \gamma$ amplitude 
by several orders of magnitude. However, even after taking this increase 
into consideration the inclusive branching ratio due to the 
$c \to u \gamma$ short distance penguin reaches only the $10^{-8}$ region, 
which is still much smaller than the effect of long distance contributions. 
Thus, in order to estimate these decays, one must concentrate on the 
treatment of the long distance dynamics involved in the $D \to V \gamma$ 
transitions. 

During the last few years several papers have appreared in which 
the $D \to V \gamma$ transitions were considered. 
In ref. \cite{BGHP} the first comprehensive 
phenomenological study of the various $D \to V \gamma$ has been presented, 
using mainly the techniques of pole diagrams and vector meson dominance. 
Other approaches include the use of the quark model 
picture and of effective Lagrangian \cite{AK,HYC0}, the use of QCD sum rules 
\cite{KSW} and the hybrid model approach, which combines heavy 
quark effective theory and chiral Lagrangian  \cite{BFO1,BFO0,FS}.
As it will be emphasized in the last section, the existing predictions of 
the various theoretical attempts are quite divergent for some of the 
$D \to V \gamma$ modes which underlines the urgent need for experimental 
data as well as for the development of a reliable model. 

It should also be mentioned that these decays offer certain opportunities to 
search for 
signals of physics beyond the standard model \cite{BFO0,BFO1,BGM}, 
although some of the proposed tests could be affected by the long 
distance contributions embodied in the $c \to u \gamma$ transition \cite{FS}. 

In the present paper we aim for a more systematic and comprehensive treatment 
of these decays than previously undertaken, employing the formalism of the 
hybrid model \cite{casone,castwo} for treating the $D \to V \gamma$ transition 
\cite{BFO0,FS}.
In Section 2, we present the details of our approach and we define the 
approximation used, in Section 3 we give the explicit form of the 
amplitudes and we conclude in Section 4 with a discussion and a 
comparative presentation of our numerical predictions.\\

{\bf 2 Model description}

\vspace{0.5cm}

We treat the radiative decays $D \to V \gamma$ as originating from the 
nonleptonic 
transition $D \to V V_0$, followed by the conversion $V_0 \to \gamma$ via the 
vector meson dominance mechanism. Although a similar scheme has been 
considered also in previous papers \cite{BGHP,BFO1,FS} there is no 
systematic treatment which includes all possible diagrams within the chosen 
approximation. In the present paper we adopt the factorization approach 
for the $D \to V V_0$ amplitude, following the formalism advanced in \cite{BSW} 
for the nonleptonic  decays of $D$, $D_s$ mesons (BSW scheme).
We shall not repeat here the arguments 
for using the factorization approximation, as these have been amply discussed
in the literature (see, e.g. \cite{BSW,AJB}). 
We are aware that the nonfactorizable contributions could also play 
a role as it may be the case for certain D \cite{KAMAL} 
and B \cite{SOARES} nonleptonic decays.
However, at the present time, before any actual measurements 
of $D \to V \gamma$ exist, we prefer to limit ourselves to 
a simple scheme, and to keep our approach as transparent as possible, 
awaiting the confrontation with experiment.
The factorization amplitude for $D \to V V_0$ in BSW scheme is calculated 
using the effective nonleptonic weak Lagrangian 
\begin{eqnarray}
\label{weak}
{\cal L}_{w}
& = & -{G_F \over \sqrt{2}} V_{uq_i}V_{cq_j}^*  
~[ a_1 ({\bar u} q_i)^{\mu}
({\bar q_j}c )_{\mu} 
%\nonumber\\
 +   a_2 ({\bar u} c)_{\mu} 
({\bar q}_j q_i)^{\mu}], 
\end{eqnarray}

\noindent
where $({\bar \psi}_1\psi_2)^\mu\equiv
{\bar \psi}_1\gamma^\mu(1-\gamma^5)\psi_2$, $q_{i,j}$ represent the fields of 
$d$ or $s$ quarks, $V_{ij}$ are the CKM matrix elements and $G_F$ is the 
Fermi constant. 
In our calculation we use $a_1 = 1.26$ and $a_2 = -0.55$ as 
found in \cite{BSW}. 

In Eq. (\ref{weak}) the quark bilinears are treated as interpolating 
fields for the appropriate mesons. In order to calculate the 
matrix elements we use as before \cite{BFO1,FS} the hybrid 
model which combines the heavy quark effective and 
chiral perturbation theory \cite{casone,castwo}. 
The relevant hadronic degrees of freedom within this 
framework are the charm pseudoscalar ($D$) and vector ($D^*$) mesons 
and the light pseudoscalar ($P$) and vector ($V$) mesons. 
In the factorization scheme ('' vacuum insertion'') which we use, 
the $D \to V V_0$ amplitude is schematically 
approximated as follows

\begin{eqnarray}
\langle V V_0 | (\bar q_iq_j)^{\mu}(\bar q_kc)_{\mu} |D\rangle &=&
\langle  V | (\bar q_iq_j)^{\mu}|0\rangle\langle V_0|(\bar q_kc)_{\mu} |D\rangle
\nonumber\\
&+&\langle  V_0 | (\bar q_iq_j)^{\mu}|0\rangle\langle V|(\bar q_kc)_{\mu} 
|D\rangle
\nonumber\\
&+&\langle  V V_0 | (\bar q_iq_j)^{\mu}|0\rangle\langle 0|(\bar q_kc)_{\mu} 
|D\rangle~,
\label{factor}
 \end{eqnarray}
where the first two terms are the spectator contributions, in the following 
denoted by $A_{Spec, \gamma}$ and $A_{Spec,V}$, respectively, and the third term 
is the weak annihilation contribution, denoted by $A_{Annih}$. 

In the three terms of Eq. (\ref{factor}), the $V_0$ meson 
($\rho^0$, $\omega$ and $\Phi$) produced in the transition is allowed to convert 
into a photon through the vector meson dominance (VDM).
The diagrams thus contributing to the amplitudes  
$A_{Spec,\gamma}$, $A_{Spec,V}$  and $A_{Annih}$ are shown in the Figs. 1a, 1b 
and 1c, respectively. However, as a result of the specific form of the strong 
Lagrangian of the heavy particles (see Eqs. (11), (19) of \cite{FS}), 
there is also direct emission of the photon from the initial D meson, as 
exhibited in diagrams (C) and (D) of Fig. 1a.

The square in each diagram of 
Fig. 1 denotes the weak transition due to the effective Lagrangian ${\cal 
L}_{w}$ (\ref{weak}). This Lagrangian contains a product of two left handed 
quark currents $(\bar \psi_1\psi_2)^{\mu}$, each denoted by a dot on Fig. 1. In 
our model, the left handed currents will be expressed in terms of the relevant 
hadronic degrees of freedom: $D$, $D^*$, $P$ and $V$. In our notation in the 
diagram ($B$), for example, the hadronic current $J_2$ creates $V$ meson, while 
the hadronic current $J_1$ annihilates $D$ and creates $V_0$ at the same time.    

In ref. \cite{FS} (FS) which also uses the hybrid model for these decays, 
most diagrams 
exhibited in Fig. 1 have already been calculated. We shall not repeat 
this calculation 
here and shall combine the results of FS with those of our systematic approach, 
to 
obtain the full $D \to V \gamma$ amplitude in the factorization approximation. 
We rely on FS as a complementary source for various basic expressions giving 
here only those formulae which are directly necessary for the calculations of 
the present paper.

The principal contribution missing in FS is due to diagram ($B$) of Fig. 1a. 
As it turns out, the inclusion of this parity - violating (PV) 
contribution, alters considerably the numerical values of the FS amplitudes 
in the PV sector and leads to the set of predictions for these decays exhibited 
in 
Table 2, 
which will be discussed in the last Section. 
The relevant expressions needed for diagram ($B$) are given below.

The weak current $J_1$ of diagram ($B$) (see Fig. 1) annihilates quark $c$ and 
creates one of the light quarks $q$ ($u$, $d$ or $s$): $J_1^{\mu}={\bar 
q}\gamma^\mu(1-\gamma^5)c$. Under chiral $SU(3)_L\times SU(3)_R$ this quark 
current transform as $({\bar 3}_L,1_R)$. At the hadronic level we impose the 
same chiral transformation and we require the current to be linear in the 
heavy meson fields $D^a$ and $D^{*a}_\mu$ \cite{BFO2,casone,castwo}

\begin{eqnarray}
\label{jqbig}
{J_1}_{a}^{\mu} = &\frac{1}{2}& i \alpha Tr [\gamma^{\mu}
(1 - \gamma_{5})H_{b}u_{ba}^{\dag}]\nonumber\\
&+& \alpha_{1}  Tr [\gamma_{5} H_{b} ({\hat \rho}^{\mu}
- {\cal V}^{\mu})_{bc} u_{ca}^{\dag}]\nonumber\\
&+&\alpha_{2} Tr[\gamma^{\mu}\gamma_{5} H_{b} v_{\alpha} 
({\hat \rho}^{\alpha}-{\cal V}^{\alpha})_{bc}u_{ca}^{\dag}]+...\;,
\end{eqnarray}

\noindent
where $\alpha=f_H\sqrt{m_H}$ and $\alpha_1$ and $\alpha_2$ are free parameters, 
which have to be determined from the experiment. 
The current (\ref{jqbig}) is the most general one in the leading $1/m_c$ order 
of HQET and next to leading order of chiral perturbation theory 
\cite{BFO2,casone,castwo}.  
Here both the heavy pseudoscalar and the heavy vector 
mesons were incorporated in a $4\times 4$ matrix $H_a$
\begin{eqnarray}
\label{defh}
H_a& = & \frac{1}{2} (1 + \!\!\not{\! v}) (P_{a\mu}^{*}
\gamma^{\mu} - P_{a} \gamma_{5})\;,
\end{eqnarray}
\noindent
where $a=1,2,3$ is the $SU(3)_V$ index of the light 
flavours, and $P_{a\mu}^*$, $P_{a}$, annihilate a 
spin $1$ and spin $0$ heavy meson $Q \bar{q}_a$ of 
velocity $v$, respectively. 
The fields $\cal V$ and $u$ incorporate the light pseudoscalars and 
are given in FS.  
The field $\hat \rho$ incorporates the light vector mesons
\begin{eqnarray}
\label{defrhohat}
{\hat \rho}_\mu & = & i {{\tilde g}_V \over \sqrt{2}} \rho_\mu\;~~,~~~~~~
\rho_\mu = \pmatrix{
{\rho^0_\mu + \omega_\mu \over \sqrt{2}} & \rho^+_\mu & K^{*+}_\mu \cr
\rho^-_\mu & {-\rho^0_\mu + \omega_\mu \over \sqrt{2}} & K^{*0}_\mu \cr
K^{*-}_\mu & {\bar K^{*0}}_\mu & \Phi_\mu \cr}. 
\end{eqnarray}

\noindent
where ${\tilde g}_V=5.9$ was fixed in the case of exact flavour symmetry (see 
e.g. 
references in FS).

The weak current $J_2$ of diagram ($B$) (see Fig. 1)  creates the final $V$ 
meson. Its matrix elements are given by \cite{FS} 
\begin{equation}
\langle V(\epsilon_V,q)| J_{2\mu} |0\rangle = \epsilon_{\mu}^{*} (q) g_V(q^2)~,
\label{j2} 
\end{equation}  
where the couplings $g_V(m_V^2)$ are measured in the leptonic decays of the 
mesons: 
$g_{\rho}(m_{\rho}^2) \simeq g_{\rho}(0)= 0.17~ GeV^2$, 
$g_{\omega}(m^2_{\omega}) \simeq g_{\omega} (0) = 0.15 ~GeV^2$, 
 $g_{\Phi}(m^2_{\Phi}) \simeq g_{\Phi} (0) = 0.24 ~GeV^2$ and 
 $g_{K^*} = (m_{K^*} /m_{\rho})g_{\rho}$.

For the calculation of the amplitude, we also need the $\gamma -  V_0$ 
interaction Lagrangian, given by the vector meson dominance like in 
\cite{BFO1,FS} 
\begin{eqnarray} 
\label{VMD}
{\cal L}_{V\gamma} = -{e\over \sqrt{2}} B_{\mu} (g_{\rho}\rho^{0\mu} + 
\frac{g_{\omega}}{3} 
\omega^{\mu} - \frac{{\sqrt 2g_{\Phi}}}{3} \Phi^{\mu})~,
\end{eqnarray}
where $B_{\mu}$ is the photon field. \\

{\bf 3 Decay amplitudes}

\vspace{0.5cm}

In order to facilitate the incorporation of the results of FS we shall 
adopt here their notation of the amplitudes, namely $A_{PC}^i$, $A_{PV}^i$ 
for the parity - conserving and parity - violating parts, where $i$ 
denotes classes of diagrams identified below. The general gauge 
invariant amplitude $D(p) \to V(p_V) \gamma(q)$ is  
\begin{eqnarray}
A(D(p)  \to V(\epsilon_{(V)},p_{(V)}) \gamma(\epsilon_{(\gamma)},q)) & = & 
e \frac{G_F}{{\sqrt 2}} V_{uq_i} V_{cq_j}^* 
\{\epsilon_{\mu \nu \alpha \beta} q^{\mu} \epsilon_{(\gamma)}^{*\nu} 
p^{\alpha}
\epsilon_{(V)}^{*\beta} A_{PC} \nonumber\\
 +  i [( \epsilon_{(V)}^{*} \cdot q)(\epsilon_{(\gamma)}^{*} \cdot p_{(V)}) &-&
(p_{(V)} \cdot q)( \epsilon_{(V)}^{*} \cdot \epsilon_{(\gamma)}^{*})] A_{PV}\},
\label{amplitude}
\end{eqnarray}
with  $A_{PC}=A_{PC}^I+A_{PC}^{II}+A_{PC}^{III}$ 
and $A_{PV}=A_{PV}^I+A_{PV}^{II}+A_{PV}^{III}$.
Now, concerning  the classification of diagrams, $A_{PC}^I$ 
will denote the contribution from diagrams ($A$) and ($C$) (see Fig. 1)
which encompass the 
$D^* \to D \gamma$ transition, while $A_{PC}^{II}$ will denote diagram ($G$)  
which contains the $P \to V \gamma$ transition. 
$A_{PC}^{III}$, $A_{PV}^{III}$ denote the 
contribution of the long distance penguins described by Fig. 1 of FS. 
On the hadronic level they are represented by the diagrams ($E$) and ($F$), 
respectively, and they vanish in the exact SU(3) flavour limit as shown in FS.  
We also define two other classes of parity violating diagrams:
$A_{PV}^I$ will include the 
bremsstrahlung-like diagrams ($D$) and ($H$), 
whereby the photon emission is due to the direct coupling to charged initial D 
or final $V$ mesons. Finally,   
$A_{PV}^{II}$, which has not been studied before, will denote the contribution 
represented by the diagram ($B$). As this contribution will be considered in 
detail below, we  present its quark level picture in Fig. 2: Fig. 2a represents 
the diagram ($B$)  for the decays proportional to $a_1$, for example 
$D_s^+\to \rho^+\gamma$, while Fig. 2b represents the decays proportional to 
$a_2$, for example $D_0\to \bar K^{*0}\gamma$.

Now we turn to the calculation of the amplitude for the diagram ($B$), while the 
remaining contributions have been studied in FS. First we parametrize the matrix 
element $\langle V_0|J_1^{\mu}|D\rangle $ in the usual way \cite{WSB1}        
 \begin{eqnarray}
\label{parhv}
&\langle& \!\!\!\!  V_0(q,\epsilon_{V0})|J_1^\mu|D(p)\rangle =
\\
&=&{2 V(Q^2)\over m_D+m_{V0}}
\epsilon^{\mu\nu\alpha\beta}\epsilon_{V0\nu}^* p_\alpha
q_\beta 
+ i 2{\epsilon^*_{V0}\cdot Q \over Q^2} m_{V0}Q^\mu ( A_3(Q^2) - A_0(Q^2))
\nonumber\\ 
&+& i(m_D+m_{V0})\biggl[\epsilon_{V0}^{\mu *} A_1(Q^2) 
-{\epsilon^*_{V0} \cdot Q\over (m_D+m_{V0})^2}(p+q)^{\mu}
A_2(Q^2)\biggr] \; \nonumber,
\end{eqnarray}
where $Q = p - q$. In fact the diagram ($B$) contributes only to form factors 
$A_0$, 
$A_1$, $A_2$ and $A_3$, which will be determined using the model described 
above. The form factor $V(Q^2)$ gets contribution from  
the diagrams ($A$) and ($C$) of Fig.1, its results are explained  in FS, 
therefore 
we will leave this contribution aside here. 
In order that these matrix elements 
be finite at $Q^2 = 0$, the form factors satisfy the relation 
\cite{WSB1}  

\begin{equation}
\label{relff}
A_3(Q^2)-{m_D+m_{V0}\over 2 m_{V0}}A_1(Q^2)+
{m_D-m_{V0}\over 2 m_{V0}}A_2(Q^2)=0\;,
\end{equation}
and $A_3 (0) = A_0(0)$. 

Using the currents (\ref{j2}) and (\ref{parhv}) and  the relation ({\ref{relff}) 
we determine the amplitude of the diagram ($B$) for $D_s^+\to \rho^+\Phi$ as an 
example:  
\begin{eqnarray}
& {\cal 
A}_{PV}&\!\!\!\!(D_s^+(p)\to\rho^+(p_{(V)},\epsilon_{(V)})\Phi(q,\epsilon
_{(\Phi)}))=  
{G_F
\over \sqrt{2}} a_1  V_{cs}^*V_{ud}
(m_D+m_{\Phi})\nonumber\\
&\times&\!\!\!\!\!\!\biggl(
\epsilon_{(\Phi)}^{*\mu}~A_1(m_V^2)-{(\epsilon_{(\Phi)}^*\cdot p_{(V)}) 
\over 
(m_D+m_{\Phi})^2}(p+q)^{\mu}A_2(m_V^2)\biggr)~g_{V}~\epsilon_{(V)\mu}^*.
%~{g_{\Phi}\over 3 m_{\Phi}^2}~
\label{aaa}
\end{eqnarray}
According to the vector meson dominance (\ref{VMD}), the amplitude for  
$D_s^+ \to \rho^+ \gamma$ is obtained, if the polarization 
$\epsilon_{(\Phi)}^{*\mu}$  
is repaced by $\epsilon_{(\gamma)}^{*\mu}e g_{\Phi}/ (3 m_{\Phi}^2)$. However, 
the amplitude for $D_s^+ \to \rho^+ \gamma$ decay should satisfy the gauge 
invariance condition.  
It was found  \cite{GP,SO} for the case of  $B \to K^* \gamma$ decay, that it 
 is useful 
to analyze the heavy meson decays into $ VV_0$ in terms 
of helicity amplitudes 
of the two final vector meson: 
${\cal A}_{++}$, ${\cal A}_{--}$ and ${\cal A}_{00}$. Thus,  
the application of gauge invariance condition to 
$D\to VV_0$ decay, with $V_0 \to \gamma$ conversion,  
means that the ${\cal A}_{00} (D\to VV_0)$ helicity amplitude  
must be discarded. Under a gauge transformation as implemented by 
$\epsilon_{(\gamma)}^{\mu} 
\to q^{\mu}$, we derive  the following general condition 
for the $D \to V V_0 \to V \gamma$ decays 
\begin{equation} 
\sum_{V^0} (m_D + m_{V^0}) \biggl[ A_1(m_V^2) - \frac{m_D^2 - m_V^2}{(m_D + 
m_{V_0})^2} A_2(m_V^2)\biggr]  =  0
\label{gaugc1}
\end{equation}
imposed for the decays of type ($B$) graphs, 
presented on Fig. 1a. Consequently, the 
${A}_{PV}^{II}$ amplitude can be expressed in terms of the form factor 
$A_1(m_V^2)$ only.
  
Now we determine the form factors $A_1(m_V^2)$ 
for the diagram ($B$) 
using the current (\ref{jqbig}) 
and parametrizing it in the form of (\ref{parhv}).  
The weak current (\ref{jqbig}) determines the form factor in the heavy quark 
limit, i.e. at the maximum momentum transfer $Q_{max}^2=(m_D-m_{V0})^2$  
\begin{eqnarray}
\label{A1}
A_1^{DV_0}(Q^2_{max} ) & = &- \frac{{\tilde g}_V}{{\sqrt 2}} 
2 \alpha_1 \frac{{\sqrt m_D}}{m_D + m_{V0}}~.
\end{eqnarray}
We assume the pole dominance behaviour of the form factors \cite{casone,castwo}
and at $Q^2 = m_V^2$ we determine 
\begin{eqnarray}
\label{A1PV}
A_1^{DV_0}(m_V^2)&  = &- {\tilde g}_V {\sqrt 2} 
\alpha_1 \frac{ {\sqrt m_D}}{m_D + m_{V0}}
\frac{1 - \frac{(m_D - m_{V0})^2}{m_{D_1^+}^2}}
{1 - \frac{(m_V)^2}{m_{D_1^+}^2}}  ,
\end{eqnarray}
where $D_{1^+}$ is the mass of the ${\bar q} c$ $J^{P}= 1^+$ 
bound state. We use the masses of ${\bar s}c$ and  ${\bar d}c$ bound states 
to be $2.53$ $ GeV$ and $2.42$ $GeV$ as in \cite{castwo}. The free parameter 
$\alpha _1$ is determined by  using the average of experimental 
$A_1(0)$ values for $D_s^+\to\Phi l\nu_l$ and $D^+\to\bar K^{*0}l\nu_l$. We 
obtain  $|\alpha_1| = 0.171$ $ GeV^{1/2}$ and use this value for the prediction 
of all $D\to V\gamma$ decay rates. 

Using the formalism described above, with Eqs.(\ref{aaa}), (\ref{gaugc1}), 
(\ref{A1PV}), we obtain $A_{PV}^{II}$ for the various decay channels.

The Cabibbo allowed decay amplitudes, which are proportional to 
the product $|V_{ud} V_{cs}^*|$, are: 

\begin{eqnarray}
A_{PV}^{II} (D^{0} \to \bar K^{*0} \gamma) & =& -a_2 
\biggl[\frac{g_{\rho} g_{K^*}}{m_{\rho}^2} (m_D  + 
m_{\rho})|A_1^{D\rho}(m_{K^*}^2)|\nonumber\\
& +&  \frac{g_{\omega} g_{K^*}}{3m_{\omega}^2} (m_D + 
m_{\omega})|A_1^{D\omega}(m_{K^*}^2)|\biggr]
~\frac{1}{m_D^2 - m_{K^*}^2}
\label{apvd0}
\end{eqnarray}

\begin{eqnarray}
A_{PV}^{II} (D^{+}_s \to \rho^{+} \gamma)&  = & a_1 ~
\frac{2g_{\Phi} g_{\rho} }{3 m_{\Phi}}(m_D + 
m_{\Phi})|A_1^{D_s\Phi}(m_{\rho}^2)|~
\frac{1} {m_{D_s}^2 - m_{\rho}^2} .
\label{apvds}
\end{eqnarray} 

 The Cabibbo suppressed 
amplitudes $A_{PV}^{II}$ proportional to the Cabibbo factor $|V_{su} V_{cs}^*|$ 
are  

\begin{eqnarray}
A_{PV}^{II} (D^{+} \to \rho^{+} \gamma)&  = &- a_1 \biggl[
\frac{g_{\rho}^2}{m_{\rho}^2}
(m_D + m_{\rho})|A_1^{D\rho}(m_{\rho}^2)| 
\nonumber\\
& - & \frac{g_{\omega}g_{\rho}}{3m_{\omega}^2}
(m_D + m_{\omega})
|A_1^{D\omega}(m_{\rho}^2)|\biggr]~{1\over m_{D}^2 - m_{\rho}^2} 
\label{apvdpcs}
\end{eqnarray}

\begin{eqnarray}
A_{PV}^{II} (D^{+}_s \to K^{*+} \gamma)&  = & a_1 
\frac{2g_{\Phi}g_{K^*}}{3m_{\Phi}}(m_{D_s} + m_{\Phi})
|A_1^{D_s \Phi}(m_{K^*}^2) | {1\over m_{D_s}^2 - m_{K^*}^2}
\label{apvdscs}
\end{eqnarray}

\begin{eqnarray}
A_{PV}^{II} (D^{0} \to \rho^{0}\gamma)&  = &
- \frac{a_2}{{\sqrt 2}}\biggl[\frac{g_{\rho}^2}{m_{\rho}^2}(m_D + 
m_{\rho})|A_1^{D\rho}(m_{\rho}^2)|\nonumber\\   
& + &\frac{g_{\omega}g_{\rho}}{3 m_{\omega}^2} (m_D + m_{\omega})
|A_1^{D\omega}(m_{\rho}^2)|\biggr]~{1\over m_D^2 - m_{\rho}^2}.
\label{apvdp0s}
\end{eqnarray}
\begin{eqnarray}
A_{PV} ^{II}(D^{0} \to \omega \gamma)&  = &
 \frac{a_2}{{\sqrt 2}}\biggl[\frac{g_{\rho}^2}{m_{\rho}^2}(m_D + 
m_{\rho})|A_1^{D\rho}(m_{\omega}^2)|\nonumber\\ 
& + &\frac{g_{\rho}g_{\omega}}{3 m_{\omega}^2} 
(m_D + m_{\omega})|A_1^{D\omega}(m_{\omega}^2)|\biggr]~{1\over m_D^2 
-m_{\omega}^2}.
\label{apvdp0so}
\end{eqnarray}
\begin{eqnarray}
A_{PV}^{II} (D^{0} \to \Phi \gamma)&  = &- a_2 
\biggl[\frac{g_{\rho} g_{\Phi} }{ m_{\rho}^2}(m_D + 
m_{\rho})|A_1^{D\rho}(m_{\Phi}^2)|\nonumber\\
& + & \frac{g_{\omega} g_{\Phi} }{ 3 m_{\omega}^2}(m_D + m_{\omega})
|A_1^{D\omega}(m_{\Phi}^2)|\biggr] ~{1\over m_{D}^2 - m_{\Phi}^2} .
\label{apvdphi}
\end{eqnarray}
For completeness, we give also the parity violating parts of the 
amplitudes for doubly suppressed decays proportional to $|V_{us} V_{cd}^*|$:

\begin{eqnarray}
A_{PV}^{II} (D^{+} \to K^{*+} \gamma)&  = & -a_1 
\biggl[\frac{g_{\rho} g_{K^*}}{m_{\rho}^2} (m_D + m_{\rho}) 
|A_1^{D\rho}(m_{K^*}^2)|\nonumber\\
& - &  \frac{g_{\omega}g_{K^*}}{3m_{\omega}^2} (m_D + m_{\omega}) 
|A_1^{DK^*}(m_{K^*}^2)|\biggr]~{1\over m_D^2 - m_{K^*}^2},
\label{apvdp}
\end{eqnarray}

\begin{eqnarray}
A_{PV}^{II} (D^{0} \to K^{*0} \gamma) & = &a_2 
\biggl[\frac{g_{\rho}g_{K^*}}{m_{\rho}^2} (m_D + m_{\rho}) 
|A_1^{D\rho}(m_{K^*}^2)|\nonumber\\
& +&  \frac{g_{\omega}g_{K^*}}{3m_{\omega}^2} (m_D + m_{\omega}) 
|A_1^{D\omega}(m_{K^*}^2)|\biggr]~{1\over m_D^2 - m_{K^*}^2} .
\label{dsapvd0}
\end{eqnarray}
\\

{\bf 4 Discussion}

\vspace{0.5cm} 

 We present numerical results for the amplitudes ${\cal A}_{PC}^I$, 
 ${\cal A}_{PC}^{II}$,  ${\cal A}_{PC}^{III}$, ${\cal A}_{PV}^I$, ${\cal 
A}_{PV}^{II}$ 
and  ${\cal A}_{PV}^{III}$ in Table \ref{tab1}, where ${\cal A}^i$ denotes 
\begin{equation}
{\cal A}_{PC(V)}^i= e {G_F \over {\sqrt 2}} V_{uq_j} V_{c q_k}^* A_{PC(V)}^i~.
\label{def}
\end{equation}
and $i$ runs over the nine decays studied.  
The amplitudes ${\cal A}_{PC}^{II}$ are calculated from Eqs. (15) - (23), 
while for the rest of the amplitudes we use the values obtained in FS. 
Although we have the values of Table 1, which encompass all the 
amplitudes arising from our factorization scheme and hybrid model, we cannot 
predict at this stage definite  values for the decay rates. This is due to 
the fact that the signs of several constants 
entering the expressions of ${\cal A}^i_{PC}$, ${\cal A}^i_{PV}$ 
(like $\lambda'$, $\lambda$, $C_{VV\Pi}$, $g_V$, $\alpha_1$, defined in FS 
and here) are not determined yet from the experimental data. 
Thus, we must contain ourselves in the present 
only to the range of values, which are obtained by assuming all 
possible relative signs for the various constants. 
We also remark that ${\cal A}^{III}_{PC}$, ${\cal A}^{III}_{PV}$ are usually 
one to two orders of magnitude smaller than the other amplitudes and will 
affect the rates only in the case of cancellations occuring among the rest of 
amplitudes. 

We calculate the branching ratios of the $D \to V \gamma$ decays 
with the help of 
\begin{equation}
\Gamma (D \to V \gamma) = 
\frac{1} {4 \pi}\biggl[ \frac{ m_D^2 - m_V^2}{ 2 m_D}\biggr]^3 ( |{\cal 
A}_{PC}|^2 + 
|{\cal A}_{PV}|^2 ) , 
\label {parw}
\end{equation}
and we present the possible range of values for the branching ratios in Table 2. 
We compare the present results, denoted by (a),  with the results 
obtained in previous approaches.
The results of the approach \cite{FS} are denoted by (b), 
the results of \cite{BGHP}  by (c), and the results of \cite{KSW} by (d) 
(we specify $a_1 = 1.26$ and  $a_2 = - 0.55$ in their formulas). 
The quark model calculation of \cite{AK} predicts the 
branching ratio for the Cabibbo allowed decays 
$BR(D^0 \to {\bar K}^{*0} \gamma) = 8.6 \cdot 10^{-6}$ 
and $BR(D^+_s \to \rho^+ \gamma) = 2.1 \cdot 10^{-5}$, which are smaller 
in comparison with the results we obtain here, as well as compared 
with the predictions of \cite{BGHP} and \cite{FS}. On the other hand, 
the calculation of \cite{HYC0}, which also uses a quark model, leads to a larger 
branching ratio $BR(D^0 \to {\bar K}^{*0} \gamma) = 1.1 \cdot 10^{-4}$, which  
 is an order of magnitude larger than obtained in \cite{AK} and 
closer to our estimate.  
We notice that the parity violating amplitudes calculated within the 
present approach have changed significantly in 
comparison with the results of FS. Overall, inspection of Table 2, 
indicates that our predictions in the present paper 
are closest to those of 
\cite{BGHP}. When measurements of a few channels will be 
available it will be possible to adjust the range of the various branching 
ratios and to make firmer predictions, hopefully 
allowing to select the best suited model.

The measurable ratio ${\cal A}_{PC}/{\cal A}_{PV}$ can 
also be used to distinguish between the various models. 
However, since at the present stage we cannot specify 
the relative signs of the various components of each of 
these amplitudes, we are unable to make any sensible statement 
about these ratios.
 
We summarize our results as follows:
we have presented a calculation of the radiative $D \to V \gamma$ decays
using a model which contains all classes of diagrams arising 
from the factorization approach for the $D \to V V^0$ amplitude, 
from which the radiative decays are obtained by use 
of vector meson dominance. 
In the calculations of 
the various matrix elements, we use a hybrid 
model \cite{casone,castwo} which 
combines heavy quark techniques with the chiral Lagrangian. 
In view of uncertainties related to the coupling constants involved, 
we can predict at this stage 
only ranges for the branching ratios of the various decay channels, which are 
given in Table 2. We emphasize that the Cabibbo allowed decays 
$D^0 \to {\bar K}^{*0} \gamma$ 
and $D^+_s \to \rho^+ \gamma$ are calculated to be fairly frequent, with 
branching ratios of a few times $10^{-4}$ and 
we expect their detection soon. Some of the Cabibbo suppressed modes, like 
$D^{+,0} \to \rho^{+,0} \gamma$ may also occur with branching ratios close to 
$10^{-4}$. 
Experimental results on these modes 
are eagerly awaited and will certainly contribute to clarify the long distance 
dynamics 
leading to these radiative decays.\\

{\it Acknowledgements:} The research of S.F. and S.P. was supported in part by 
the Ministry of Science of the Republic of Slovenia. 
The research of P.S. was supported in part by Fund for 
Promotion of Research at the Technion. \\

\vspace{0.7cm}

{\bf Figure Captions}

\vspace{0.5cm}

{\bf Fig. 1.} Skeleton diagrams of various contributions to the long distance 
decay $D \to V \gamma$. The spectator diagrams of type $A_{Spec,\gamma}$ (see 
Eq. (\ref{factor}) in the text) are shown in Fig. 1a, the spectator diagrams of 
type $A_{Spec,V}$ are shown in Fig. 1b and the weak annihilation diagrams 
 $A_{Annih}$ are shown in Fig. 1c.  The square in each diagram 
denotes the weak transition due to the effective Lagrangian ${\cal 
L}_{w}$ (\ref{weak}). This Lagrangian contains a product of two left handed 
quark currents $(\bar \psi_1\psi_2)^{\mu}$, each denoted by a dot. 
 Different diagrams are denoted by 
($A$) - ($H$). Their contributions to the amplitudes ${\cal A}_{PC}^i$ and 
${\cal A}_{PV}^i$ are specified in the text.

{\bf Fig. 2.} The quark level picture of the diagram ($B$) of Fig. 1 
(amplitude $A_{PV}^{II}$):
Fig. 2a represents the decays proportional to $a_1$, for example $D_s^+\to 
\rho^+\gamma$, while Fig. 2b represents the decays proportional to $a_2$, for 
example $D_0\to \bar K^{*0}\gamma$.   

\newpage
\begin{table}[h]
\begin{center}
\begin{tabular}{|c||c|c|c|c|c|c|}
\hline
$D\to V \gamma$ & $|{\cal A}_{PC}^{I}|$  
& $|{\cal A}_{PC}^{II}| $ 
& $|{\cal A}_{PC}^{III}| $
& $|{\cal A}_{PV}^I| $ & ${\bf |{\cal A}_{PV}^{II}|}$
& $|{\cal A}_{PV}^{III}| $ \\
\hline \hline
$ D^0 \to {\bar K}^{*0} \gamma$ &$ 6.4$ & $6.2$ & $0$ & $0$ & $5.5$ & $0$  \\
\hline
$ D_s^+ \to \rho^+ \gamma$ &$ 1.4$ & $7.3$ & $0$ & $7.4$ & $4.3$ & $0$  \\
\hline
\hline
$ D^0 \to \rho^{0} \gamma$ &$ 0.82$ & $1.0$ & $0.02$& $0$ & $0.71$ & $0.03$ \\
\hline
$ D^0 \to \omega \gamma$ &$ 0.73$ & $1.07$ & $0.02$& $0$ & $0.63$ & $0.03$  \\
\hline
$ D^0 \to \Phi \gamma$ &$ 1.8$ & $1.34$ & $0$& $0$ & $1.8$ & $0$  \\
\hline
$ D^+ \to \rho^+ \gamma$ &$ 0.59$ & $1.3$ & $0.02$& $1.6$ & $1.3$ & $0.03$  \\
\hline
$ D_s^+ \to K^{*+ }\gamma$ &$ 0.41$ & $2.3$ & $0.02$& $2.1$ & $1.2$ & $0.04$  \\
\hline
\hline 
$ D^+ \to K^{*+} \gamma$ &$ 0.16$ & $0.42$ &$ 0$ &  $0.43$ & $ 0.37$ & $0$  \\
\hline
$ D^0 \to K^{*0} \gamma$ &$ 0.33$ & $0.32$ &$ 0$ &  $0$ & $ 0.28$  & $0$\\
\hline
\end{tabular}
\caption{ The parity conserving and parity violating amplitudes for 
charm meson decays in units $10^{-8}$ $GeV^{-1}$. The amplitudes ${\cal 
A}^i_{PC,PV}$ get contributions from different diagrams in Fig. 1: (A) and (C) 
contribute to ${\cal A}_{PC}^I$, (D) and (H) contribute to ${\cal A}_{PV}^{I}$, 
(G), (B), (E) and (F) contribute to ${\cal A}_{PC}^{II}$, ${\cal A}_{PV}^{II}$, 
${\cal A}_{PC}^{III}$ and ${\cal A}_{PV}^{III}$, respectively. 
The amplitudes ${\cal 
A}_{PC}^I$, ${\cal 
A}_{PC}^{II}$,  ${\cal A}_{PC}^{III}$, ${\cal A}_{PV}^I$ and  ${\cal 
A}_{PV}^{III}$ were calculated in \cite{FS}, while the amplitude ${\cal 
A}_{PV}^{II}$ represents the additional contribution calculated here.  The first 
two decays are 
Cabibbo allowed, while the last two are doubly Cabibbo suppressed. }
\label{tab1}
\end{center}
\end{table}
\newpage
\begin{table}[h]
\begin{center}
\begin{tabular}{|c||c|c| c| c|}
\hline
$D\to V \gamma$ & $BR(a)\times 10^{5}$ & $BR(b)\times 10^{5}$ & 
$BR(c)\times 10^{5}$ &$BR(d)\times 10^{5}$ \\
\hline \hline
$ D^0 \to {\bar K}^{*0} \gamma$ &$ (6-36) $ & $(10^{-2}-30)$ & $ (7-12) $ 
& $0.18$ \\
\hline
$ D_s^+ \to \rho^+ \gamma$ &$ (20-80)$ & $ (34-50)$ & $ (6-38) $ & $ 4.4$ \\
\hline
\hline
$ D^0 \to \rho^{0} \gamma$&$ (0.1-1)$ & $(0.02 - 1) $ & $(0.1 - 0.5)$
& $ 0.38 $\\
\hline
$ D^0 \to \omega \gamma$ &$ (0.1 - 0.9)$ & $ (0.02 - 0.8)$ & 
$ \simeq 0.2$ & $ - $ \\  
\hline
$ D^0 \to \Phi \gamma$ &$ (0.4 - 1.9 ) $ & $ (0.04 - 1.6)$ & 
$(0.1-3.4)$ & $ - $\\
\hline
$ D^+ \to \rho^+ \gamma$ &$ (0.4 -6.3)$ & $(1.8 - 4.1)$ & $ (2 - 6)$ & $ 0.43$  
\\ 
\hline
$ D_s^+ \to K^{*+ }\gamma$ &$(1.2 - 5.1)$ & $ (2.1 - 3.2)$ & $ (0.8-3)$ & $-$\\
\hline
\hline 
$ D^+ \to K^{*+} \gamma$ &$ (0.03- 0.44)$ & $(0.12 - 0.25)$ & $ 0.1- 0.3$ & $ 
-$\\
\hline
$ D^0 \to K^{*0} \gamma$ &$ (0.03 - 0.2) $ & $ (10^{-5} - 0.08)$ & $ \simeq 0.01$ 
& 
$- $    \\ 
\hline
\end{tabular}
\caption{ The branching ratios for $D \to V \gamma$ 
decays. The first column  (a) contains the results  of the present approach.  
The next three columns present the results of Ref. \cite{FS} (b),  
Ref. \cite{BGHP} (c), 
and  Ref. \cite{KSW} (d). The first two decays are Cabibbo allowed, 
while the last two are doubly Cabibbo suppressed.
}\label{tab2}
\end{center}
\end{table}

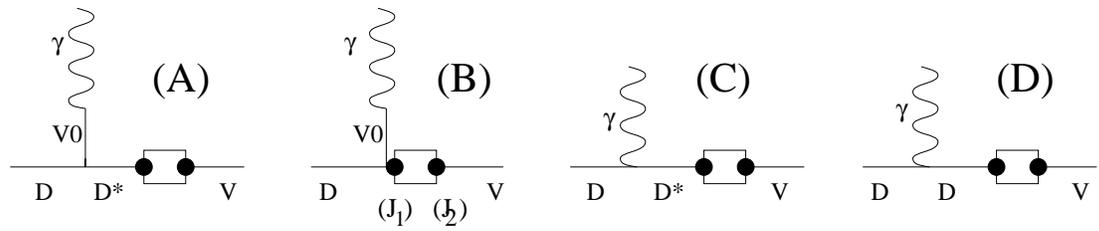
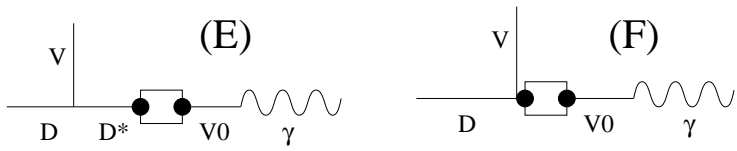
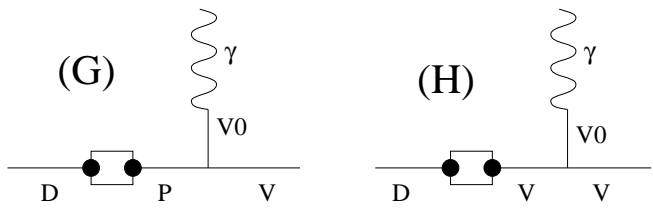
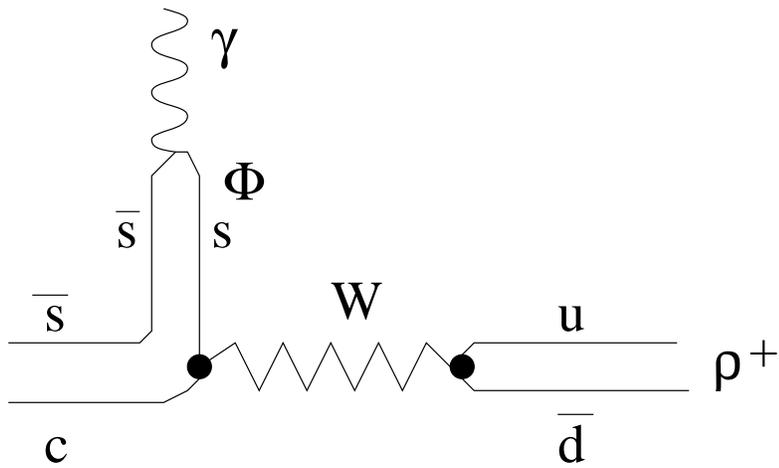
\end{document}